\documentclass{iaus}
\usepackage{graphicx}

\title[The large-scale structure in the CDFS ] 
{The large-scale structure\\ in the Chandra Deep Field South}

\author[]   
{Dario Trevese$^1$,
 Fabrizio Fiore$^2$,  \\Enrico Piconcelli$^2$, Marco Castellano$^2$, Laura Pentericci$^2$,\\ Piero Ranalli$^{3,4}$ \and Andrea Comastri$^4$} 

\affiliation{$^1$ Sapienza Universit\`a di Roma, P.le Aldo Moro 5, 001865 Roma, Italy \\
email: dario.trevese@roma1.infn.it\\
$^2$ INAF-Osservatorio Astronomico di Roma, via Frascati 33, 00040 Monteporzio (RM), Italy\\
$^3$ Dipartimento di Astronomia, Universit\`a di Bologna, via Ranzani 1, 40127 Bologna, Italy \\
$^4$ INAF-Osservatorio Astronomico di Bologna, via Ranzani 1, 40127 Bologna, Italy }
\pubyear{2011}
\volume{277}  
\pagerange{119--126}
\jname{Tracing the Ancestry of Galaxies (on the land of our ancestors)}
\editors{C. Carignan, K.C. Freeman \& F. Combes, eds.}
\begin{document}

\maketitle

\begin{abstract}
We have studied the large scale distribution of matter  in the Chandra Deep Field South on the basis of photometric redshifts and we have  identified several over-densities between redshift 0.6 and 2.3. We analyse  two of these structures using the deepest X-ray observations ever obtained: 4 Ms with the Chandra satellite and 2.5 Ms with XMM-Newton. We set a very faint upper limit on the X-ray luminosity of a structure at redshift 1.6,  and we find an extended X-ray emission from a structure at redshift 0.96 of which we can estimate the gas temperature and make a comparison with the  scaling relations between the X-ray luminosity and mass or temperature of high redshift galaxy clusters.

\end{abstract}

\firstsection 
\section{Introduction}

Being the largest and most massive bound structures in the Universe, galaxy clusters provide  the most biased environment for galaxy evolution. Thus, they are ideal  laboratories for studying the physical processes responsible for galaxy formation and evolution. With optical selection, it is relatively easy to obtain large statistical samples of clusters. The main disadvantage is that these samples are seriously affected by projection effects. Only expensive spectroscopic campaigns can confirm overdensities in 3 dimensions.  The X-ray detection is more robust against line-of-sight contamination, but it has a lower efficiency and higher observational cost as compared to optical surveys.  Studying the X-ray properties of optically selected clusters is essential to understand the selection effect. On one hand, extremely deep X-ray surveys with Chandra and XMM-Newton, make the detection  of the highest redshift cluster possible, on the other hand they are also extending the minimum luminosity, i.e. the least massive structures, to which X-ray clusters can be detected and analyzed at intermediate redshifts. Galaxy groups and clusters with  $kT \lesssim$3 keV are more likely to display the effects of non gravitational energy into the intracluster medium (ICM) than hotter, more massive clusters. 

In previous studies  we have proposed a "(2+1) Dimensions" method, based on photometric redshifts (\cite{trev+07}), to detect overdensities in the large scale distribution of galaxies, down to the faintest  limits attainable nowadays with optical wide-band photometric surveys. A comprehensive study of the Chandra Deep Field South (CDFS) allowed us to detect  12 overdensities with redshifts in the range  $0.6<z<2.3$  (\cite{cast+07}, \cite{sali+09}).

We present preliminary results of the analysis of the deepest X-ray  images available to date, from the 4 Ms Chandra observations of the CDFS recently made public by the Chandra X-ray Center, and 2.5 Ms XMM-Newton observations (\cite{coma+11}), restricted to two over-densities, with spectroscopic redshifts  z=0.96 and    z=1.6 respectively. 

In the following we adopt  $H_0$=70 km s$^{-1}$Mpc$^{-1}$, $\Omega_m$=0.3, $\Omega_{\Lambda}$=0.7

\section{Observations}

\begin{figure}[b]
\begin{center}
 \includegraphics[width=5.0in]{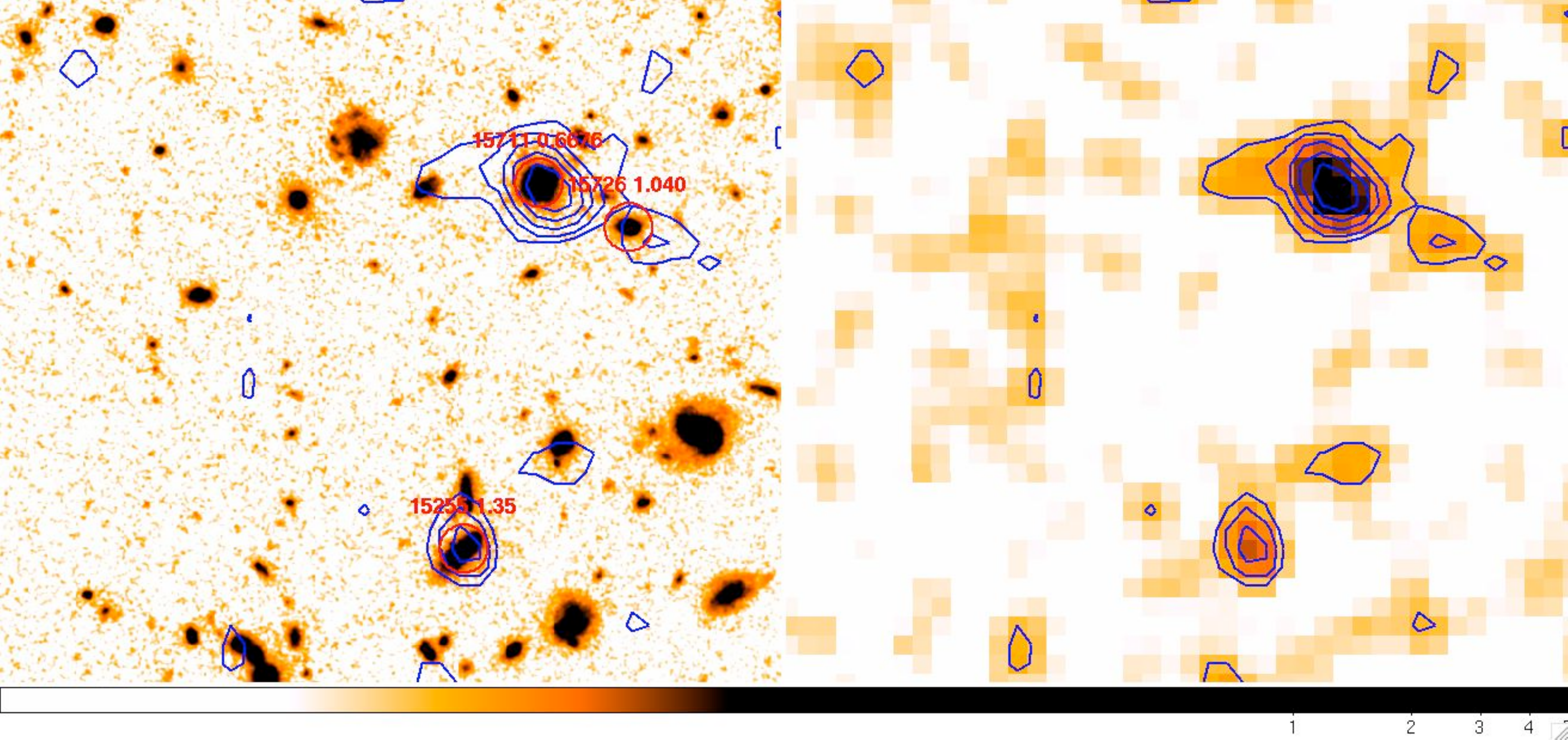} 
 \caption{{\it Left}: Hubble Space Telescope WFC3  H band image. The spectroscopic redshift and  MUSIC catalogue identification of the X-ray sources are indicated. {\it Right}: Chandra 4 Ms image of the same field in the 0.5-2 keV band. The X-ray isophotes are shown on both images.}
   \label{fig1}
\end{center}
\end{figure}

{\underline{\it The over-density at z=1.61}}
All of the three  X-ray sources detected in the area of the over-density at $z=1.61$, correspond to objects identified in the GOODS-MUSIC catalogue (\cite{graz+06}), with redshifts 0.67, 1.04 and 1.35, thus not belonging to the structure at z=1.61 (Fig. \ref{fig1}). Eliminating these sources, we obtain a $3\sigma$ upper limit to the flux  $F(0.5-2keV)<2.5\times10^{-16}$erg/cm2/s which corresponds to a bolometric luminosity
$L<10^{43}$ erg/s, for $T$= 3keV.
\cite{kurk+09} provided new spectroscopic redshifts,  analysed the velocity distribution  and estimated  a virial mass $M_{vir}=9.0 \times 10^{13} M_{\odot}$. Assuming a typical density profile we can convert the mass to $M_{500}=6.4\times10^{13}$ $M_{\odot}$ , where $M_{500}$ is the mass within a sphere with a mean inner density 500 times the critical density $\rho_c$.

\begin{figure}[t]
\vspace*{0.5 cm}
\begin{center} 
\includegraphics[width=5.3in]{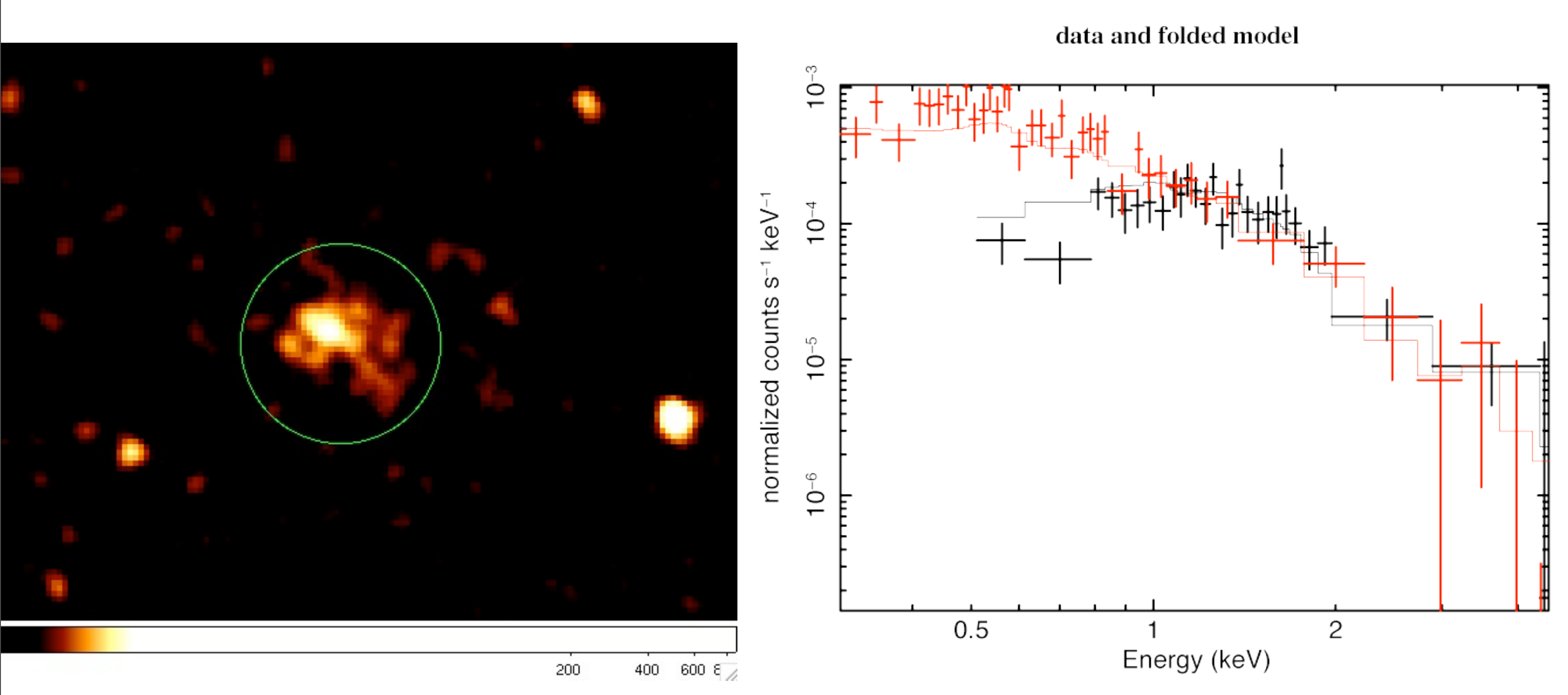} 
 \caption{{\it Left}: Chandra 4 Ms image showing the extended emission from the region of the overdensity at z=0.96. The  circle indicates the aperture adopted to extract the X-ray spectrum.
{\it Right}: simultaneous fit of Chandra and XMM-Newton  spectra of the extended source  at z=0.96,  with a MEKAL model with T=2.6 keV and assumed abundance $Z/Z_{\odot}$=0.3. The XMM spectrum (red, in the electronic version) extends to lower energy.}
\label{fig 2}
\end{center}
\end{figure}

{\underline{\it The over-density at z=0.96}}
An extended source is detected in the region of the overdensity at z=0.96 (Fig. \ref{fig 2}). The peak of the X-ray emission coincides with a normal galaxy at z=0.96 with no emission lines (\cite{szok+04}). Notice that this implies a revision of the earlier classification of the X-ray source as Type 2 AGN.

Considering the extended character of the emission and the absence of AGN detections in the cluster area, we fitted simultaneously the Chandra and XMM-Newton X-ray spectra with a MEKAL model, with an abundance $Z/Z_{\odot}$=0.3, obtaining a temperature $T=2.6^{+0.5}_{-0.3}$  keV  (1$\sigma$ error) (Fig. 2).

From 11 redshifts available in the cluster area, the velocity dispersion is $\sigma_v$=420 Km/s, which would corresponds to a virial mass: 

$M_{vir}=7.3\times10^{13} M_{\odot}$ or  to $M_{500}=5.2\times10^{13} M_{\odot}$,
assuming complete relaxation and following \cite{kurk+09}.
\eject

The M-T scaling relation obtained by \cite{pope+05} for clusters with mean redshift $z\sim0$:   
$M_{500}=2.89 \times 10^{13}  T^{1.59} M_{\odot}$ would give, instead $M_{500}=1.3\times  10^{14} M_{\odot}$, possibly suggesting an evolution of the M-T relation from $z\sim0.1$ to $z\sim1$.

\vspace*{0.5 cm}

\section{Results}

\begin{figure}[t]
\begin{center}
 \includegraphics[width=5.0in]{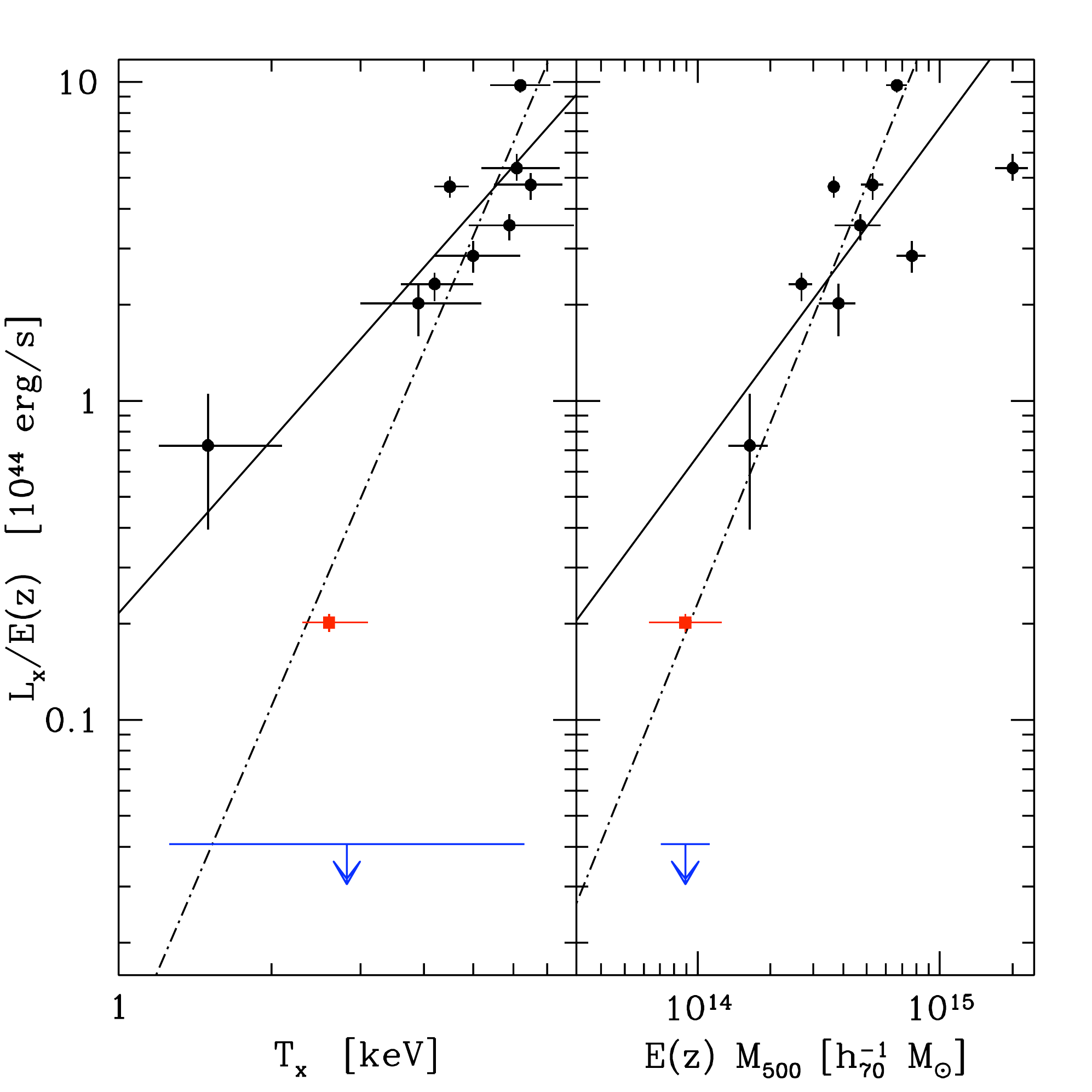} 
 \caption{ Bolometric X-ray luminosity versus temperature ({\it left}) and versus mass ({\it right}), where $E(z)=[\Omega_m (1+z)^3+ \Omega_{\Lambda}]^{1/2}$. Filled circles: clusters in the range $0.6 < z < 1.2$ from \cite{hick+08}; filled square: structure at (z=0.96); arrow: structure at (z=1.61). Continuous line: fit to the \cite{hick+08} points; dashed line: scaling relation of \cite{etto+04} for an X-ray selected sample with $0.4<z<1.3$. }
   \label{fig 3}
\end{center}
\end{figure}

In figure \ref{fig 3}  we compare our results with the $L_X-T$ and $L_X-M$ scaling relations obtained: i) by 
\cite{hick+08} for high redshift clusters with $ 0.6<z<1.2$, optically selected by the Red-Sequence Cluster Survey (RCS) (\cite{glad+05}); ii) by \cite{etto+04} for X-ray selected clusters in the range $ 0.4<z<1.3$.
 
\begin{itemize}
\item{For the structure at z=1.61 we indicate our $3\sigma$ upper limit on $L_X$, while for the unknown temperature we indicate the range 1.2-6 keV. In both diagrams the structure is underluminous with respect to the scaling relations. This suggests it is a forming cluster which did not reach full relaxation.}
\item{The structure at z=0.96  is consistent with the steeper scaling relations of \cite{etto+04} and not with those of \cite{hick+08}, despite it was selected in the optical band.}
\item{ For the other structures we identified in the  CDFS (\cite{cast+07}, \cite{sali+09}) the analysis of the deepest X-ray data existing to date is in progress.}
\item{The present result, if confirmed, would strengthen the evidence of a strong redshift evolution of the scaling relation, of the type found by Ettori et al. 2004.}
\item{The detection of very low X-ray luminosity objects among optically selected structures could be related to the fact that our "(2+1) D" method (\cite{trev+07}), unlike the RCS, is independent of the presence of a red sequence, which tends to be less evident at high redshift.}
\end{itemize}

\eject

\end{document}